\documentclass[12pt]{article}
\usepackage{graphicx}
\usepackage{authblk}

\usepackage{amsmath}
\usepackage{afterpage}
\usepackage{geometry}
\usepackage{makecell}
\usepackage{booktabs}
\usepackage{multirow}
\usepackage{array}
\usepackage{hyperref}


\title{Predicting food taste with bound-driven optimization}
\author[1]{Pagkratis Tagkopoulos}
\author[1]{Dimitris Sfondilis}
\author[1,2,3,*]{Ilias Tagkopoulos}
\author[1,3,4,*]{Tarek Zohdi}
\affil[1]{Process Integration and Predictive Analytics, PIPA LLC, Davis, USA}
\affil[2]{Department of Computer Science and Genome Center, University of California, Davis, USA}
\affil[3]{USDA AI Institute for Food Systems (AIFS), Davis, CA, USA}
\affil[4]{Department of Mechanical Engineering, University of California, Berkeley, USA}
\affil[*]{Corresponding Authors: itagkopoulos@ucdavis.edu, zohdi@berkeley.edu}
\date{\vspace{-5ex}}

\begin{document}
\maketitle

\begin{abstract}
The prediction of sensory attributes from ingredient-level formulations is an emerging challenge at the intersection of food science and artificial intelligence. We address the fundamental question of whether the taste of a food can be predicted from its ingredients by treating recipes as composite materials. We apply Hashin--Shtrikman (HS) and Reuss--Voigt (RV) bounds, techniques originally developed for elastic moduli, to predict five taste dimensions (sweetness, sourness, bitterness, umami, saltiness) on a curated dataset of 70 recipes decomposed into 209 ingredient-level taste references with trained-panel ground truth. The bounds provided an  additive baseline but systematically under-predict perceived taste: 77\% of actual taste values exceeded the HS upper bound, with the exceedance rate ranging from 26\% (bitterness) to 97\% (saltiness). We traced this gap to specific processing chemistry (Maillard reactions, caramelization, evaporative concentration, protein hydrolysis, and nucleotide synergy) and introduced a hybrid model that augments the HS baseline with eight chemistry-proxy features encoding these mechanisms. Our results show that our interpretable hybrid model eliminates the systematic bias and reduces mean absolute error by 27--62\% for sweetness, sourness, umami, and saltiness while using only 10 interpretable features, achieving performance comparable to a black-box Lasso regression on 115 per-ingredient features. We further demonstrate constrained inverse design via Differential Evolution, recovering ingredient formulations that match target taste profiles subject to compositional bounds. Our work demonstrates how key chemical processes during food preparation can inform and augment physics-based and machine learning models, providing a quantitative fingerprint of processing chemistry's contribution to taste perception and paving the way for model-driven food formulation with targeted sensory characteristics.
\end{abstract}


\section{Introduction}
Human taste---the perception of chemical stimuli by taste receptor cells---is typically described through five basic modalities: sweet, sour, salty, bitter, and umami \cite{chandrashekar2006receptors}. Each corresponds to distinct classes of tastants: sugars for sweetness, acids for sourness, sodium salts for saltiness, plant alkaloids for bitterness, and glutamate derivatives for umami. The human mouth contains thousands of taste receptor cells distributed across taste buds, each capable of detecting and integrating multiple taste stimuli \cite{chaudhari2010cell}. This sensory input gives rise to a highly complex and dynamic taste space, which becomes even more intricate when combined with olfaction, texture, temperature, and auditory cues---factors that together constitute the broader perception of flavor \cite{gunning2025systematic}. Flavor perception is thus inherently multi-sensory, shaped by chemical and cognitive processes \cite{mcgee2004food}.

A central question in computational gastronomy is whether the taste profile of a recipe can be predicted from its ingredient composition and their proportions. Conversely, the inverse problem asks which ingredients and in what proportions should be combined to achieve a desired sensory profile. Both problems have been the subject of recent research, although most work focuses on single molecules rather than composite foods. Models may predict a single taste modality from molecular structure \cite{elhadad2025decoding}, predict the full taste profile via multi-objective optimization \cite{androutsos2024predicting}, or fine-tune large language models on flavor databases \cite{zimmermann2025chemical}. Prediction of food taste as a composite whole---from ingredient-level properties to recipe-level perception---remains an open problem with far-reaching applications in food design \cite{kuhl2025ai, gunning2025systematic, schreurs2024predicting, ahn2011flavor, ghasemi2023food}.

\begin{figure}[p]
    \centering
    \includegraphics[width=\textwidth]{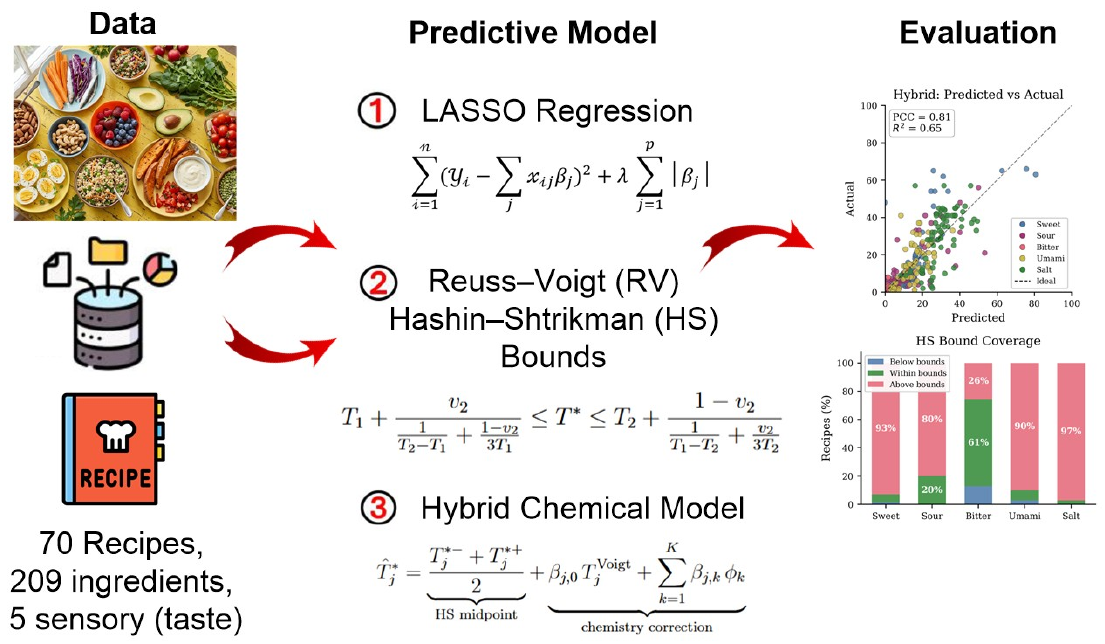}
    \vspace{-15em}
    \caption{Overview. Left: 70 recipes decomposed into 209 ingredients with five taste dimensions. Center: three modeling approaches---Lasso regression, Reuss--Voigt (RV) and Hashin--Shtrikman (HS) composite-material bounds, and a hybrid chemistry-corrected model. Right, top: hybrid model predicted vs.\ actual taste scores (PCC $= 0.80$, $R^2 = 0.63$). Right, bottom: HS bound coverage by taste dimension, showing that 80--97\% of actual values exceed the upper bound for sweet, sour, umami, and salt, while 61\% of bitterness values fall within bounds (floor effect).}
    \label{fig:fig1}
\end{figure}

In this study, we ask: \textit{can composite-material theory bracket the taste of a food?} We apply Hashin--Shtrikman and Reuss--Voigt bounds \cite{hashin1963variational}---originally developed to estimate effective elastic properties of multiphase materials---to predict recipe taste from ingredient taste scores and mass fractions. We quantify where the analogy holds and where it breaks, trace the failures to specific food-chemistry mechanisms, and introduce a hybrid model that combines the physics-informed baseline with chemistry-informed corrections. We evaluate all models against a curated dataset of 70 recipes with trained-panel ground truth on five taste dimensions (\textbf{Figure~\ref{fig:fig1}}), and demonstrate constrained inverse design via evolutionary optimization.

\section{Methods}

\subsection{Dataset}
Ground-truth taste scores (sweetness, sourness, bitterness, umami, saltiness, 0--100 Spectrum\texttrademark{} scale) for 627 Dutch food items were drawn from the trained-panel SVT database of Mars et~al.\ \cite{mars2020taste}. From this database, 126 single-ingredient items served directly as ingredient-level taste references, and 83 additional ingredients were scored from published sources (USDA FoodData Central, NEVO 2021, peer-reviewed sensory studies). Seventy multi-ingredient food items were decomposed into constituent ingredients with mass fractions assigned via a four-tier evidence hierarchy (label-declared, NEVO recipe documentation, food-science literature, arithmetic residual), with all fractions summing to $1.000 \pm 0.001$ per recipe. Each recipe carries a composition confidence rating (High: 22, Moderate: 42, Low: 6). Fat sensation, the sixth SVT dimension, was excluded because it reflects textural properties not consistently modeled by compositional approaches. The complete data pipeline is documented in the Supplementary Information, and all data are provided in Supplementary Data File~1.

\subsection{Forward prediction: composite-material bounds}

We treat each recipe as a heterogeneous $N$-phase composite, where ingredient $i$ has taste score $T_i$ and mass fraction $v_i$ ($\sum_i v_i = 1$). Each of the five taste dimensions is modeled independently.

\smallskip
\noindent\textit{Reuss--Voigt (RV) bounds.}
The Voigt bound \cite{voigt1928lehrbuch} (arithmetic weighted average) and Reuss bound \cite{reuss1929berechnung} (harmonic weighted average) give the widest theoretical brackets on the effective property $T^*$:
\begin{equation}
T^*_\text{Reuss} = \left[\sum_{i=1}^{N} \frac{v_i}{T_i}\right]^{-1} \leq\; T^* \;\leq \sum_{i=1}^{N} v_i\, T_i = T^*_\text{Voigt}
\label{eq:rv}
\end{equation}
When any $T_i = 0$ (e.g., sugar has Sour $= 0$), we regularize $T_i \to \max(T_i, \varepsilon)$ with $\varepsilon = 0.01$.

\smallskip
\noindent\textit{Hashin--Shtrikman (HS) bounds.}
For $N$ isotropic phases, the HS bounds provide tighter limits than RV \cite{hashin1963variational, zohdi2012introduction}. Defining the auxiliary function
\begin{equation}
A(T_0) = \left[\sum_{i=1}^{N} \frac{v_i}{T_i + 2\,T_0}\right]^{-1} - 2\,T_0
\label{eq:hs-aux}
\end{equation}
the multi-phase HS bounds are
\begin{equation}
T^{*-} = A(T_{\min}) \;\leq\; T^* \;\leq\; A(T_{\max}) = T^{*+}
\label{eq:hs}
\end{equation}
where $T_{\min} = \min_i T_i$ and $T_{\max} = \max_i T_i$. The midpoint $(T^{*-} + T^{*+})/2$ serves as the point estimate. The factor ``2'' in Eq.~\eqref{eq:hs-aux} arises from the spherical inclusion geometry in three dimensions, and it equals $(d-1)$ in $d$ dimensions. We adopt $d=3$ following the original derivation. We emphasize that this is a phenomenological analogy: taste intensity is not a physical modulus, and $d$ serves as a shape parameter controlling bound tightness rather than a literal spatial dimension. The robustness analysis (Supplementary Section~S9.1) confirms that the processing gap persists across $d \in \{2, 3, 5, 10, 50\}$, supporting the utility of the analogy despite its heuristic nature. For $N=2$, Eq.~\eqref{eq:hs} reduces to the classical two-phase result \cite{hashin1963variational}. Derivations are given in Supplementary Section~S3.

\subsection{Chemistry-informed hybrid model}
\label{sec:hybrid}

The HS bounds capture additive mixing but not the processing chemistry that amplifies taste during cooking: Maillard reactions \cite{hodge1953maillard} generate umami-active peptides and sweet furanones from amino acids and sugars \cite{belitz2009food}. Caramelization creates sweet and bitter compounds from thermal sugar decomposition. Evaporative water loss concentrates dissolved tastants (particularly NaCl). Protein hydrolysis releases free glutamate, and nucleotide synergy (IMP $\times$ glutamate) amplifies umami perception non-linearly \cite{breslin2013evolutionary}.

We augment the HS baseline with a learned correction over $K = 8$ chemistry-proxy features $\phi_k$ that encode these mechanisms from the ingredient list alone (no process parameters required):
\begin{equation}
\hat{T}_j^* = \underbrace{\frac{T_j^{*-} + T_j^{*+}}{2}}_{\text{HS midpoint}} + \underbrace{\beta_{j,0}\,T_j^{\text{Voigt}} + \sum_{k=1}^{K} \beta_{j,k}\,\phi_k}_{\text{chemistry correction}}
\label{eq:hybrid}
\end{equation}
where $T_j^{\text{Voigt}}$ is the Voigt bound for taste dimension $j$, and the features $\phi_k$ are: protein fraction, sugar fraction, their product (Maillard potential), salt fraction, water fraction, concentration factor $1/(1 - v_{\text{water}})$, allium fraction (onion, garlic, leek), and fermented-ingredient fraction (soy sauce, cheese, vinegar, mustard). The coefficients $\beta_{j,k}$ are learned by Lasso regression with the HS midpoint and RV bound as input features alongside the chemistry proxies. Feature definitions and the ingredient categories used to compute each $\phi_k$ are detailed in Supplementary Table~S2.

\subsection{Machine learning baseline}

Lasso regression \cite{tibshirani1996lasso} was used as a data-driven baseline, with the regularization parameter $\alpha$ selected per taste dimension via leave-one-out cross-validation (LOOCV) over 30 log-spaced values from $10^{-3}$ to $10^{1}$. Two Lasso configurations were evaluated: (i)~a 5-feature model using the RV-weighted taste vector as input, and (ii)~a 115-feature model using per-ingredient mass fractions. All features were standardized within each LOOCV fold.

\subsection{Inverse design via Differential Evolution}

To address the inverse problem---determining ingredient mass fractions that achieve a desired taste profile---we embed the hybrid forward model (Eq.~\ref{eq:hybrid}) within a Differential Evolution (DE) optimizer. The hybrid model is preferred over HS or RV for inverse design because its lower prediction error and near-zero bias (Table~\ref{tab:coverage}) ensure that optimized formulations will taste approximately as predicted. For $N$ ingredients with property vector $P_i = \{S_i, Z_i, B_i, L_i, M_i, v_i\}$, the objective function minimizes the weighted relative error
\begin{equation}
\Pi = \sum_{j=1}^{5} W_j \frac{|T_j^{*,\text{desired}} - T_j^*|}{|T_j^*|}
\label{eq:inverse}
\end{equation}
subject to $\sum_i v_i = 1$ and user-specified bounds $v_i^{\min} \leq v_i \leq v_i^{\max}$, where $W_j$ are per-dimension importance weights. DE \cite{storn1997differential} was implemented using \texttt{scipy.optimize.differential\_evolution} \cite{virtanen2020scipy} with population size~15, crossover probability~0.8, mutation factor $F \in [0.5, 1.0]$, the \texttt{best1bin} strategy, and 500 maximum iterations. The full DE formulation and implementation details are given in Supplementary Section~S4.

\section{Results}

\subsection{Composite-material bounds provide a floor, not a bracket}
The RV and HS models both show positive correlations with actual taste (aggregate PCC $= 0.65$--$0.66$) but with massive systematic under-prediction (\textbf{Figure~\ref{fig:fig1}}, evaluation panels; \textbf{Figure~\ref{fig:fig4}}A--B). Table~\ref{tab:coverage} quantifies the bound coverage: across all 70 recipes and 5 taste dimensions, 77\% of actual values exceed the HS upper bound. The exceedance is dimension-specific: saltiness (97\%), sweetness (93\%), and umami (90\%) are almost always above bounds, sourness is above in 80\% of cases, while bitterness is the only dimension where a majority (62\%) of values fall within bounds.

The HS midpoint and RV (Voigt) bound are highly correlated ($r = 0.94$), confirming that the bounds are tight and the HS tightening adds limited practical value over a simple weighted average for this application. The bitterness result requires careful interpretation: with a ground-truth mean of 1.6 and 96\% of recipes scoring $\leq 3$ on the 0--100 scale, the ``within-bounds'' result reflects a floor effect rather than model accuracy (a constant prediction of 2.0 achieves MAE $= 1.06$, only marginally worse than the model's 0.88).

\subsection{The processing gap maps to specific chemistry}

The systematic exceedance of the HS upper bound traces to identifiable food-chemistry mechanisms \cite{belitz2009food, breslin2013evolutionary, nolden2017perceptual}. Table~\ref{tab:chemistry} summarizes, for each taste dimension, the dominant processes, the molecular species generated, and their effect on perceived taste. Detailed mechanistic descriptions are given in Supplementary Section~S5.

\textbf{Saltiness} exhibits the largest systematic bias (mean under-prediction of 24.2 points) and the highest exceedance rate (97\%). The dominant mechanism is physical: NaCl does not volatilize, so as water evaporates during cooking, dissolved salt concentrates in the remaining liquid. A recipe starting with 1\% NaCl by mass and 40\% water that cooks down to 20\% water approximately doubles its effective salt concentration---an amplification invisible to the ingredient-level model, which sees only the pre-cooking mass fraction. A secondary factor is umami--salt receptor synergy: free glutamate and nucleotides (IMP, GMP) released during cooking enhance saltiness perception at the T1R receptor level by up to 1.5$\times$ without additional NaCl \cite{breslin2013evolutionary}.

\textbf{Umami} (90\% above bounds, bias $-$12.7) is under-predicted because the dominant umami-active molecules are \textit{created} during cooking, not present in raw ingredients. Protein hydrolysis at 60--100$^\circ$C cleaves peptide bonds in meat, fish, dairy, and legumes, releasing free glutamic acid---the primary umami tastant. The Maillard reaction between amino acids and reducing sugars generates umami-active peptides in the 1--5~kDa range and kokumi-enhancing $\gamma$-glutamyl peptides that stimulate T1R1/T1R3 receptors \cite{belitz2009food}. Most importantly, nucleotide synergy creates a fundamentally non-linear effect: IMP (from meat) and GMP (from mushrooms, yeast) multiply glutamate's perceived umami intensity by up to 8$\times$ \cite{yamaguchi1971basic, breslin2013evolutionary}---an interaction that no additive model can represent. Finally, cooking alliums (onion, garlic, leek) converts their sharp thiosulfinates into umami-active $\gamma$-glutamyl peptides, explaining why allium fraction emerges as a strong predictor in the hybrid model.

\textbf{Sweetness} (93\% above bounds, bias $-$13.0) is amplified by multiple thermal pathways. Caramelization---the pyrolysis of sugars above 160$^\circ$C---generates novel sweet-tasting compounds including furanones, maltol (toasty/bread aroma), diacetyl (butterscotch), and hydroxymethylfurfural (HMF) \cite{belitz2009food}. These compounds are absent from raw ingredients and contribute perceived sweetness beyond the sugar mass fraction. The Maillard reaction adds further sweet volatiles (furanones, lactones) through amino acid--sugar condensation. Water evaporation concentrates dissolved sugars, and fat carriers (butter, oil, cream) solubilize Maillard volatiles that enhance sweetness cross-modally via retronasal olfaction.

\textbf{Sourness} (80\% above bounds, bias $-$8.4) increases through acid concentration as water evaporates, carrying citric, malic, and lactic acids to higher effective concentrations. Fermented ingredients (sauerkraut, yoghurt, vinegar, aged cheese) contribute pre-formed organic acids whose levels may exceed those captured by the SVT panel scores for the unfermented precursors. Maillard intermediates (Amadori rearrangement products) are weakly acidic, contributing additional pH depression during cooking.

\noindent \textbf{Bitterness} is the only dimension where actual values predominantly fall within bounds (62\%). However, this reflects opposing forces---Maillard bitterness creation versus sweetness-mediated suppression \cite{calvino1993perception}---operating near the perceptual floor (mean $= 1.6$, 96\% of recipes $\leq 3$ on the 0--100 scale) rather than genuine model accuracy. High-temperature Maillard products (alkylpyridines, melanoidins $>$3~kDa) are bitter, but most recipes in our corpus contain sugar or other sweet ingredients that suppress bitter perception at the receptor level \cite{keast2004modification}. The net effect is near-zero change from the additive baseline, producing apparent ``within-bounds'' predictions that are trivially correct: a constant prediction of 2.0 for all recipes achieves MAE $= 1.06$, only marginally worse than the model's 0.88.

\subsection{Hybrid model eliminates the systematic bias}

The hybrid model (Eq.~\ref{eq:hybrid}), trained and evaluated via LOOCV, eliminates the systematic bias across all dimensions (Table~\ref{tab:coverage}, rightmost columns) and reduces average MAE from 14.7 to 7.3 (50\% reduction, excluding the uninformative bitterness dimension). Predicted-vs-actual scatterplots for all four models are shown in \textbf{Figure~\ref{fig:fig4}}A--D, with per-dimension error distributions in \textbf{Figure~\ref{fig:fig4}}E. Table~\ref{tab:models} compares all models head-to-head. Mean sensory profiles for all methods are compared via radar plots in Supplementary Figure~S1.

The hybrid model (10 features) matches the 5-feature Lasso in aggregate accuracy while providing interpretable, chemistry-grounded coefficients. The 115-feature per-ingredient Lasso, despite having access to the full ingredient composition, performs \textit{worse} on average due to overfitting (notably, umami MAE degrades from 6.5 to 8.7 with 115 features). The Lasso coefficients of the hybrid model are interpretable: saltiness correction is driven by salt fraction and allium content, umami by water fraction (encoding protein hydrolysis in soups and stews) and allium fraction, sweetness by sugar fraction (caramelization potential). Full coefficients are reported in Supplementary Table~S3.

\subsection{Ingredient and recipe clusters}
\begin{figure}[p]
    \centering
    \includegraphics[width=\textwidth]{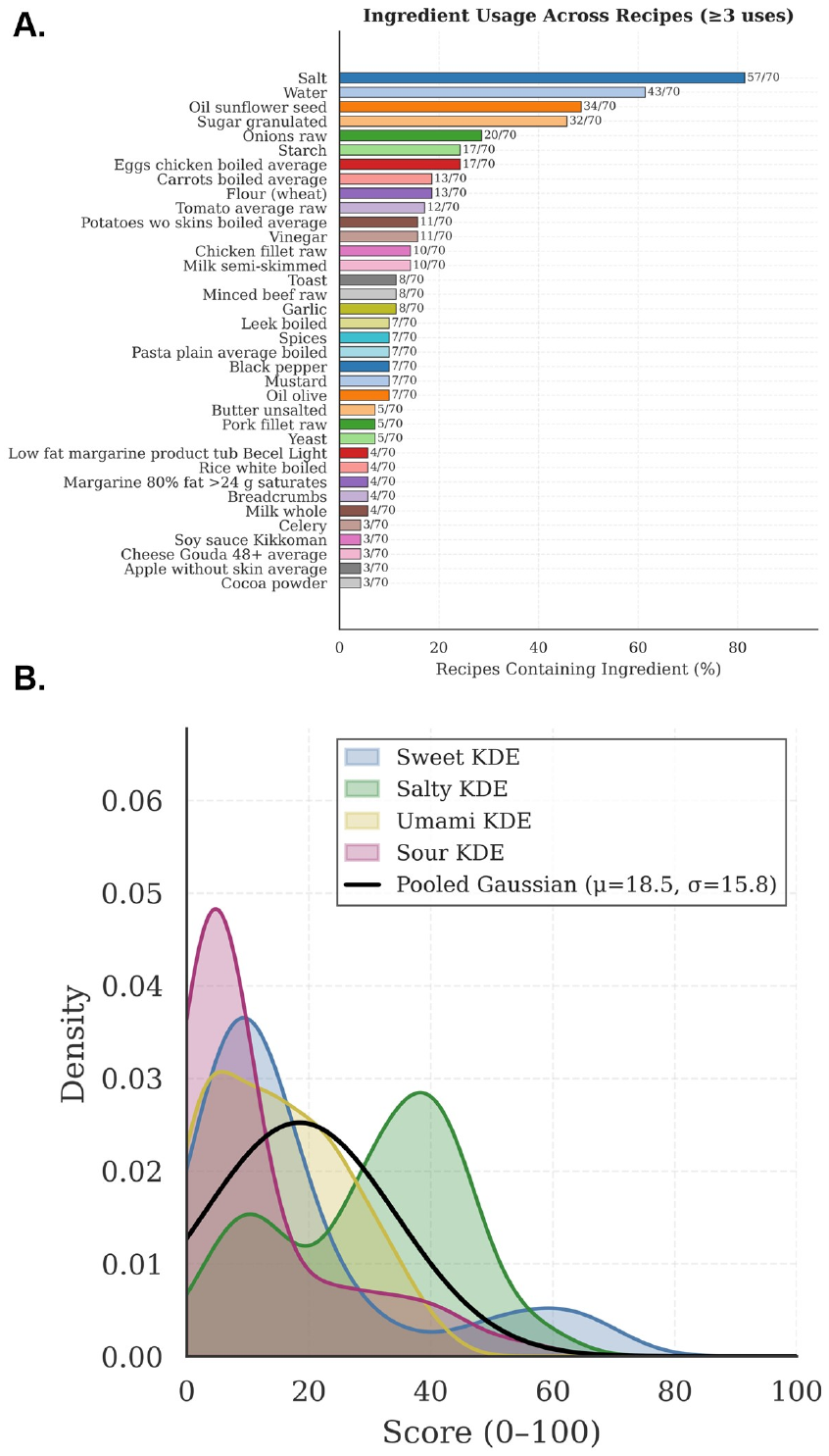}
    \caption{Dataset characteristics. (A)~Ingredient usage frequency across 70 recipes (ingredients with $\geq 3$ occurrences shown, with count out of 70). Salt, water, sunflower oil, and sugar are the most common. (B)~Kernel density estimates of recipe-level taste score distributions (0--100 Spectrum scale) for sweet, sour, umami, and salty, with pooled Gaussian overlay ($\mu = 18.5$, $\sigma = 15.8$). Bitterness is omitted due to floor effect (96\% of recipes score $\leq 3$).}
    \label{fig:fig2}
\end{figure}

\textbf{Figure~\ref{fig:fig2}} shows the distribution of ingredients (A) and taste attributes (B) across all 70 recipes. A t-SNE embedding \cite{vandermaaten2008tsne} (perplexity 12) followed by HDBSCAN density-based clustering \cite{mcinnes2017hdbscan} (min\_cluster\_size $= 7$, min\_samples $= 2$) identified five recipe clusters with a silhouette score of 0.53, outperforming $k$-means at all tested $K$ values (max silhouette 0.50 at $K = 11$). The clusters correspond to interpretable formulation categories: savory mains (C1, 16 recipes), umami-rich composed dishes (C2, 8), sweet-sour products (C3, 9), mild/low-intensity foods (C4, 16), and tomato-based dishes (C5, 7), with 14 noise points (\textbf{Figure~\ref{fig:fig3}}). Cluster assignments are provided in Supplementary Data File~1.

\subsection{Inverse design: three reformulation case studies}

We embedded the hybrid forward model within the DE optimizer to address three practically motivated reformulation problems drawn from the SVT dataset. In each case, the optimizer adjusts ingredient mass fractions to achieve a target taste profile subject to compositional constraints, with all fractions summing to 1.0. Table~\ref{tab:inverse} summarizes the results.

\textbf{Case 1 (pea soup, salt reduction).} With the hybrid forward model predicting salt at 30.5 (ground truth 44, vs.\ the HS model's 2.8), the optimizer reduced predicted saltiness from 30.5 to 25.5 ($-$16\%) by eliminating NaCl ($-$0.010) and replacing peas (24\% $\to$ 2\%) with pork ($+$0.182) and oil ($+$0.234). The increase in pork raises protein fraction, which the hybrid model's chemistry correction associates with umami preservation through protein hydrolysis, explaining why umami is maintained at 20.6 despite the dramatic compositional shift. The optimizer compensates for the lost pea volume with ingredients that do not carry salt.

\textbf{Case 2 (chocolate-hazelnut spread, sugar reduction).} The hybrid model predicts sweetness at 66.9 (ground truth 63), far more realistic than the HS prediction of 23.2. Constraining sugar to $\leq$35\% (from 50\%) with realistic ingredient bounds, the optimizer increased hazelnuts (13\% $\to$ 23\%) and milk (9\% $\to$ 20\%) while reducing oil (18\% $\to$ 10\%). Despite this compensation, predicted sweetness fell from 66.9 to 53.7---a 20\% loss---and bitterness rose from 5.0 to 6.9 due to the increased cocoa-to-sugar ratio. This quantifies a real reformulation trade-off: reducing sugar by 30\% inevitably shifts the sweet--bitter balance, and the hybrid model makes the magnitude of this shift explicit before any manufacturing trial.

\textbf{Case 3 (ketchup, umami boost).} The hybrid model predicts umami at 17.3 (ground truth 22) and sweetness at 28.1 (ground truth 28)---far more realistic than the HS predictions of 3.3 and 9.4. Targeting umami $+$3 and sweetness $-$3 with sugar capped at 5\%, the optimizer raised tomato from 60\% to 76\% and eliminated water and starch, while reducing sugar from 15\% to 10\%. The tomato concentration increases predicted umami from 17.3 to 20.3 ($+$17\%) because tomato carries the highest umami score among the ketchup ingredients, and the hybrid model's allium and fermented-ingredient corrections amplify this effect. Sweetness decreased from 28.1 to 25.1 ($-$11\%) as intended. This demonstrates multi-objective trade-off navigation: the optimizer identifies tomato concentration as the most efficient umami lever while sugar reduction provides the sweetness offset.

These three cases span savory (soup), sweet (confection), and condiment categories, demonstrating the generality of the inverse design framework. The choice of forward model matters: HS and RV systematically under-predict taste by 12--24 points on salt and umami (Table~\ref{tab:coverage}), so recipes optimized with those models would taste far saltier and more umami-rich than predicted. The hybrid model, with its near-zero bias, produces designs whose predicted taste profiles are trustworthy. The optimizer's reasoning remains transparent: each ingredient change traces to its taste-score contribution and chemistry-proxy features, grounding the reformulation in the mechanisms described in Section~3.2.

\section{Discussion}

This study demonstrates that composite-material bounds, applied here for the first time to food taste prediction, provide a useful lower envelope, but not an accurate bracket: the RV weighted average sets a physics-informed floor for perceived taste, while the HS tightening offers only marginal improvement ($r = 0.94$ between HS midpoint and RV). The central finding is that 77\% of actual taste values exceed the HS upper bound, and the per-dimension exceedance rates (Salt 97\%, Sweet 93\%, Umami 90\%, Sour 80\%, Bitter 26\%) constitute a quantitative fingerprint of processing chemistry's contribution to taste beyond additive ingredient mixing. As such, the gap between bounds and actual taste values is the measurement of how much Maillard reactions, caramelization, evaporative concentration, and protein hydrolysis contribute to perceived taste.

The hybrid model eliminates this systematic bias by augmenting the HS baseline with eight chemistry-proxy features, achieving performance comparable to a black-box Lasso while using 10$\times$ fewer features and providing interpretable, domain-grounded coefficients. There is no significant difference between hybrid and Lasso~5D errors (Wilcoxon signed-rank test, $p = 0.81$), and 10-fold cross-validation across five random splits yielded stable performance (MAE $= 7.4 \pm 0.2$). The HS bound exceedance pattern is robust to the dimensionality parameter $d$: varying $d$ from 2 to 50 changes the fraction of values above the upper bound only marginally (75--79\%), confirming that the processing gap is not an artifact of the $d = 3$ choice. Performance stratified by composition confidence shows that high-confidence recipes (HS MAE 11.7, Hybrid MAE 8.0) and moderate-confidence recipes (HS MAE 16.2, Hybrid MAE 6.8) both benefit from the hybrid correction, while the six low-confidence recipes show slightly higher hybrid error (MAE 8.4), consistent with noisier mass-fraction estimates propagating into predictions. The model's practical advantage is that the correction features are computable from the ingredient list alone, and thus no process parameters (temperature, time, pH) are required, because the features encode the \textit{potential} for processing chemistry rather than the process itself.

Several limitations constrain the present analysis. First, the sample size of 70 recipes, while a contribution by itself, limits both the statistical power and the external generalizability of our models. With 10 features in the hybrid model and 70 training samples, the ratio of observations to parameters is modest, and LOO-CV may produce optimistic estimates on out-of-distribution recipes; the high-dimensional Lasso (115 features) overfits, as evidenced by umami MAE degrading from 6.5 to 8.7. Moreover, all models were trained and evaluated on the same SVT corpus via LOO-CV, with no external validation on an independent sensory database. Prospective panel testing of the inverse-design reformulations would further strengthen confidence in generalizability. Second, the chemistry-proxy features rely on keyword-based ingredient classification that cannot distinguish between items with vastly different chemical profiles (e.g., fresh mozzarella vs.\ aged Gouda both match ``Cheese,'' despite a $\sim$10-fold difference in free glutamate content). Replacing these heuristics with quantitative nutrient-composition features from NEVO or USDA---free glutamic acid, reducing sugar content, and organic acid concentration per 100\,g---would provide more principled and granular proxies. Third, the models do not capture perceptual cross-modal interactions such as sweetness suppressing bitterness or fat enhancing sweetness, nor the transformative effects of cooking temperature and duration \cite{le2023dynamic, green2010taste}. Fourth, the bitterness dimension is uninformative in this corpus (mean 1.6, SD 1.4 on a 0--100 scale) and should not be interpreted as evidence for or against the composite-material analogy. Fifth, the HS and RV bounds assume isotropic mixing of phases, yet real food matrices are spatially heterogeneous, and the mass fractions in our dataset carry residual uncertainty from non-label-declared components (see Supplementary Information). Finally, the inverse design case studies use the hybrid forward model, which eliminates systematic bias but still carries per-recipe prediction error (MAE 7.3). The optimized formulations should therefore be interpreted as informed guidance for reformulation, with panel validation recommended before production.

Future work should pursue four directions. First, expanding the dataset to include validated recipes from independent sensory databases, and prospectively validating at least one inverse-design reformulation with a trained panel, would establish external generalizability. Second, keyword-based features can be replaced with nutrient-derived proxies (free amino acids, reducing sugars, organic acids) for a deeper and more accurate chemistry encoding. Third, incorporating process parameters to model temperature-dependent Maillard kinetics and evaporation rates would add to the predictive performance of the models. Fourth, the introduction of non-linear interaction terms, particularly nucleotide and glutamate synergy for umami, will help capture the amplification effects that linear models fundamentally cannot represent.

\section*{Data Availability}
The SVT sensory database is available at DOI: 10.17026/dans-xst-ughm. Supplementary Data File~1 containing the full recipe decomposition, model predictions, clustering assignments, and inverse design results is provided with this paper.

\section*{Code Availability}
All code for bound computation, hybrid model training \cite{pedregosa2011scikit}, inverse design, clustering, and figure generation is available as a \href{https://github.com/Pagkratios/Predicting-food-taste-from-ingredients}{github repository}. A portal that allows food profiling based on food composition will be available upon publication of the paper.

\section*{Acknowledgments}
We would like to thank the data science and multi-physics teams in PIPA LLC, and the Tagkopoulos lab in UC Davis for the helpful discussions and early prototyping of these models. 

\section*{Use of GenAI Statement}
During the preparation of this work the authors used Claude in order to check for any inconsistencies, revise, improve readability, and reformat the manuscript. After using this tool/service, the authors reviewed and edited the content as needed and take full responsibility for the content of the publication.

\section*{Competing Interests}
P.T., D.S., and I.T. work or have financial interests in PIPA LLC, an AI company in Food, Nutrition, and Health.

\bibliographystyle{unsrt}
\bibliography{bibliography}


\clearpage

\clearpage

\clearpage
\begin{figure}[p]
    \centering
    \includegraphics[width=\textwidth]{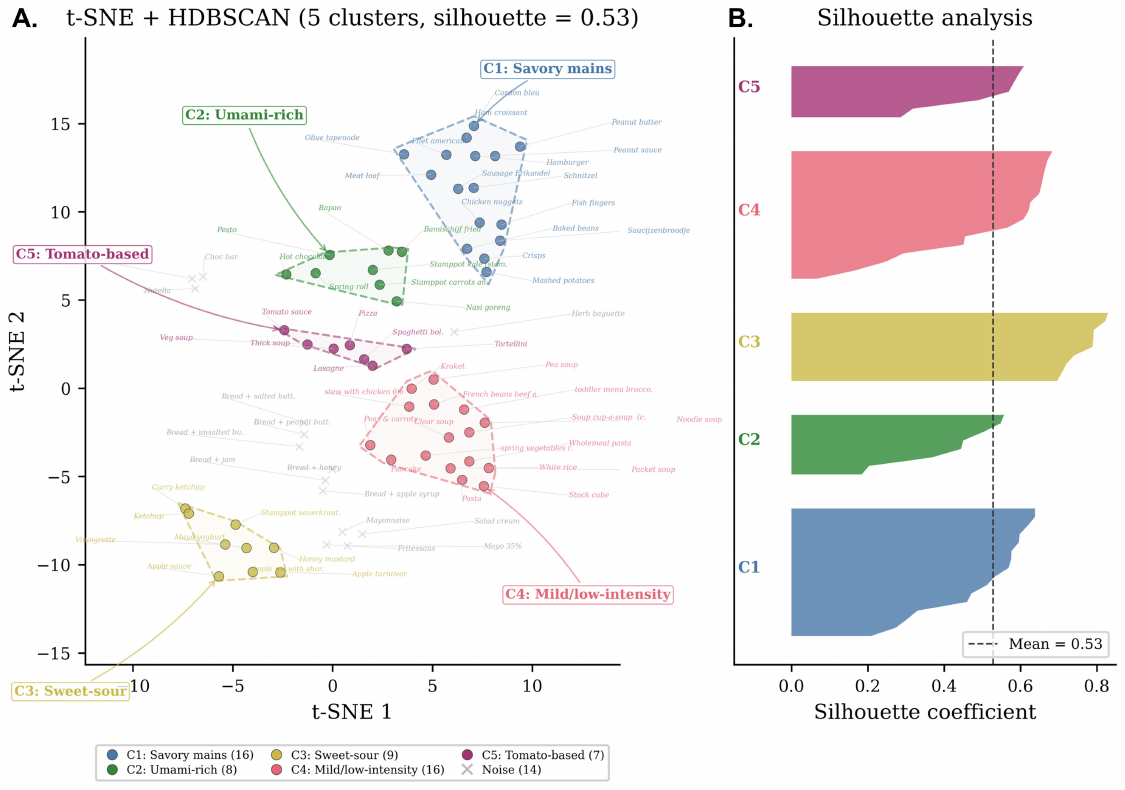}
    \vspace{-15em}
    \caption{Recipe clustering. (A)~t-SNE embedding of 70 recipes (RV-weighted taste vectors, standardized) with HDBSCAN density-based clustering (min\_cluster\_size $= 7$, min\_samples $= 2$). Five clusters were identified: C1 savory mains (16 recipes), C2 umami-rich dishes (8), C3 sweet-sour products (9), C4 mild/low-intensity foods (16), and C5 tomato-based dishes (7), with 14 noise points. Dashed boundaries show convex hulls. All recipe names are labeled. (B)~Per-cluster silhouette coefficients (mean $= 0.53$), exceeding the best $k$-means silhouette (0.50 at $K = 11$).}
    \label{fig:fig3}
\end{figure}

\clearpage
\begin{figure}[p]
    \centering
    \makebox[\textwidth][c]{\includegraphics[width=1\textwidth]{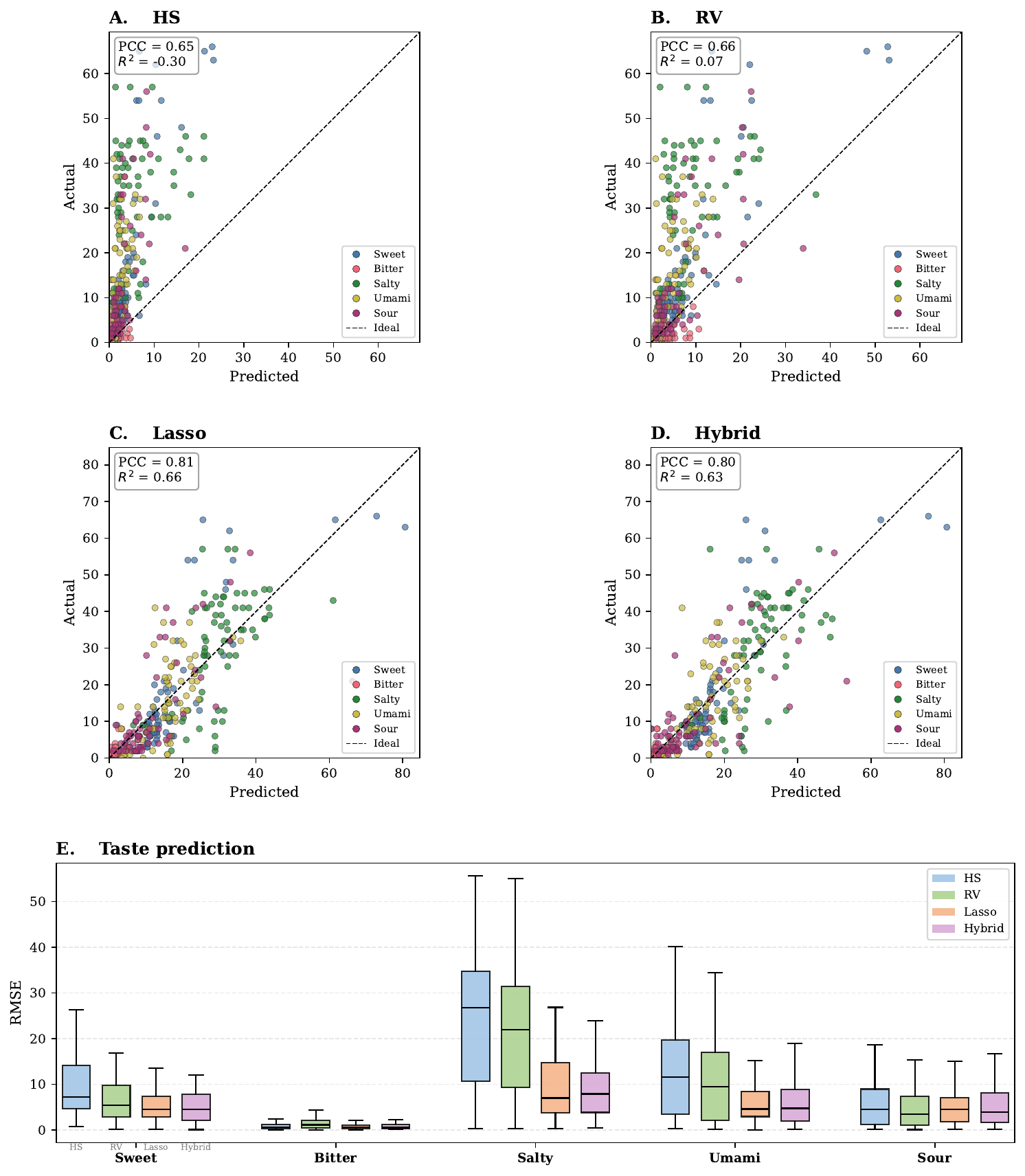}}
    \vspace{-1em}
    \caption{Model evaluation. (A)~HS bound predicted vs.\ actual taste scores (PCC $= 0.67$). (B)~RV bound predicted vs.\ actual (PCC $= 0.69$). (C)~Lasso regression predicted vs.\ actual (PCC $= 0.85$, $R^2 = 0.72$). (D)~Hybrid model predicted vs.\ actual (PCC $= 0.80$, $R^2 = 0.63$). Panels A--B use the same axis range as C--D but show the systematic under-prediction that compresses all points to the lower-left. (E)~Error distributions (RMSE) by taste dimension for all four methods, showing that HS and RV have large errors for salty and sweet while Lasso and Hybrid substantially reduce errors across all dimensions.}
    \label{fig:fig4}
\end{figure}


\clearpage
\begin{table}[t!]
\centering
\caption{HS bound coverage and model performance across taste dimensions (LOO cross-validation, $N = 70$ recipes). Avg(4D) excludes bitterness (floor effect, see text).}
\label{tab:coverage}
\small
\begin{tabular}{l rr rrr rrr rrr}
\toprule
& \multicolumn{2}{c}{\textbf{Ground truth}} & \multicolumn{3}{c}{\textbf{Bound coverage}} & \multicolumn{3}{c}{\textbf{HS midpoint}} & \multicolumn{3}{c}{\textbf{Hybrid (10 feat.)}} \\
\cmidrule(lr){2-3} \cmidrule(lr){4-6} \cmidrule(lr){7-9} \cmidrule(lr){10-12}
\textbf{Taste} & \textbf{Mean} & \textbf{SD} & \textbf{Below} & \textbf{In} & \textbf{Above} & \textbf{MAE} & \textbf{PCC} & \textbf{Bias} & \textbf{MAE} & \textbf{PCC} & \textbf{Bias} \\
\midrule
Sweet   & 17.7 & 17.6 & 1\% & 6\%  & 93\% & 13.0 & .83 & $-$13.0 & 7.6 & .70 & $-$0.5 \\
Sour    & 11.4 & 12.9 & 0\% & 20\% & 80\% &  8.6 & .63 & $-$8.6  & 6.3 & .74 &    0.4 \\
Bitter  &  1.6 &  1.4 & 13\%& 61\% & 26\% &  0.9 & .31 & $-$0.0  & 0.9 & .14$^\dagger$ &    0.0 \\
Umami   & 15.0 & 10.9 & 3\% & 7\%  & 90\% & 12.8 & .55 & $-$12.7 & 6.1 & .63 &    0.0 \\
Salt    & 29.8 & 14.6 & 0\% & 3\%  & 97\% & 24.2 & .38 & $-$24.2 & 9.3 & .59 &    0.0 \\
\midrule
Avg(4D) &      &      &     &      &      & 14.7 & .60 & $-$14.7 & 7.3 & .67 &    0.0 \\
\bottomrule
\end{tabular}
\vspace{2pt}

{\footnotesize $^\dagger$Bitter PCC is unreliable due to floor effect (96\% of values $\leq 3$ on a 0--100 scale).}
\end{table}

\clearpage
\begin{table}[t!]
\centering
\caption{Processing chemistry driving HS bound exceedance by taste dimension. ``\% above'' indicates the fraction of recipe--taste pairs exceeding the HS upper bound.}
\label{tab:chemistry}
\footnotesize
\setlength{\tabcolsep}{5pt}
\renewcommand{\arraystretch}{1.05}
\begin{tabular}{p{1.8cm} p{2.5cm} p{2.8cm} p{5.2cm}}
\toprule
\textbf{Taste} \newline (\% above) & \textbf{Mechanism} & \textbf{Key molecules} & \textbf{Effect} \\
\midrule
\textbf{Salt} \newline (97\%)
 & Evaporative conc. & NaCl (non-volatile) & Water loss during cooking concentrates dissolved salt (dominant mechanism) \\[3pt]
 & Umami--salt synergy & Glutamate, IMP, GMP & Nucleotides amplify saltiness at receptor level ($\sim$1.5$\times$) \\
\midrule
\textbf{Sweet} \newline (93\%)
 & Caramelization & Furanones, maltol, HMF & Sugar pyrolysis ($>$160$^\circ$C) creates sweet compounds \\[3pt]
 & Maillard reaction & Sweet furanones, lactones & Amino acid--sugar condensation yields sweet volatiles \\[3pt]
 & Water loss, fat synergy & Dissolved sugars, fat-soluble volatiles & Concentration plus cross-modal enhancement via retronasal olfaction \\
\midrule
\textbf{Umami} \newline (90\%)
 & Protein hydrolysis & Free glutamic acid & Heating (60--100$^\circ$C) cleaves peptide bonds, releasing umami amino acids \\[3pt]
 & Maillard reaction & Umami peptides (1--5 kDa), $\gamma$-glutamyl peptides & Novel tastants absent from raw ingredients \\[3pt]
 & Nucleotide synergy & IMP $\times$ glutamate & Non-linear amplification up to 8$\times$ \\[3pt]
 & Allium caramelization & $\gamma$-Glutamyl peptides & Onion/garlic cooking $\to$ umami-active compounds \\
\midrule
\textbf{Sour} \newline (80\%)
 & Acid concentration & Citric, malic, lactic acids & Water evaporation concentrates organic acids \\[3pt]
 & Fermentation & Lactic, acetic acid & Pre-formed acids in aged/fermented ingredients \\
\midrule
\textbf{Bitter} \newline (26\%)
 & Maillard ($>$150$^\circ$C) & Alkylpyridines, melanoidins & Bitter compounds at high temperature \\[3pt]
 & Sweetness suppression & Sweet compounds in recipe & Cross-modal: sweetness suppresses bitter perception \\
\bottomrule
\end{tabular}
\end{table}

\clearpage
\begin{table}[t!]
\centering
\caption{Model comparison (LOO cross-validation, $N = 70$ recipes, average over sweet, sour, umami, salt).}
\label{tab:models}
\small
\begin{tabular}{lcccc}
\toprule
\textbf{Model} & \textbf{Feat.} & \textbf{MAE} & \textbf{PCC} & \textbf{Interpretable} \\
\midrule
HS midpoint           & 0   & 14.7 & 0.60 & Yes \\
RV (weighted avg)    & 0   & 11.9 & 0.63 & Yes \\
Lasso (5D RV)        & 5   & 7.3  & 0.67 & Partial \\
\textbf{Hybrid HS + chem.} & \textbf{10} & \textbf{7.3} & \textbf{0.67} & \textbf{Yes} \\
Lasso (per-ingredient)& 115 & 7.5  & 0.64 & No \\
\bottomrule
\end{tabular}
\end{table}

\clearpage
\begin{table}[t!]
\centering
\caption{Inverse design results for three reformulation scenarios. ``Orig'' and ``Opt'' are hybrid-model-predicted taste scores before and after optimization. Target is the desired profile. Key ingredient changes are listed with their chemistry-grounded rationale.}
\label{tab:inverse}
\small
\begin{tabular}{p{3.8cm} p{3.8cm} p{5.5cm}}
\toprule
\textbf{Case 1: Pea soup} \newline Salt reduction \newline (RP14, 12 ingredients) &
\textbf{Case 2: Chocolate spread} \newline Sugar reduction \newline (RP55, 7 ingredients) &
\textbf{Case 3: Tomato ketchup} \newline Umami boost \newline (RP68, 7 ingredients) \\
\midrule
\textit{Constraint:} NaCl $\leq$ 0.3\% \newline \textit{Objective:} reduce salt by 5 pts, preserve umami &
\textit{Constraint:} sugar $\leq$ 35\% \newline \textit{Objective:} maintain sweetness &
\textit{Constraint:} sugar $\leq$ 5\% \newline \textit{Objective:} umami $+$3, sweet $-$3 \\
\midrule
Pork: $+$0.182 (9\% $\to$ 27\%) \newline Oil: $+$0.234 (1\% $\to$ 25\%) \newline Peas: $-$0.226 (24\% $\to$ 2\%) \newline Salt: $-$0.010 &
Hazelnuts: $+$0.101 (13\% $\to$ 23\%) \newline Milk: $+$0.110 (9\% $\to$ 20\%) \newline Sugar: $-$0.142 (50\% $\to$ 36\%) \newline Oil: $-$0.081 &
Tomato: $+$0.159 (60\% $\to$ 76\%) \newline Sugar: $-$0.051 (15\% $\to$ 10\%) \newline Water: $-$0.078 \newline Starch: $-$0.038 \\
\midrule
Salt: 30.5 $\to$ 25.5 ($-$16\%) \newline Umami: 20.6 $\to$ 20.6 \newline Sweet: 14.4 $\to$ 14.4 &
Sweet: 66.9 $\to$ 53.7 ($-$20\%) \newline Bitter: 5.0 $\to$ 6.9 ($+$38\%) \newline Umami: 4.2 $\to$ 4.2 &
Umami: 17.3 $\to$ 20.3 ($+$17\%) \newline Sweet: 28.1 $\to$ 25.1 ($-$11\%) \newline Salt: 29.8 $\to$ 29.8 \\
\bottomrule
\end{tabular}
\end{table}

\end{document}